\documentclass[journal=iecred,manuscript=article]{achemso}

\usepackage[T1]{fontenc}
\usepackage{csquotes}
\usepackage{amsmath}
\usepackage{amssymb}
\usepackage{natbib}
\usepackage{graphicx}
\graphicspath{ {./image/} }

\author{Benjamin J. Landrum}
\email{blandrum@alumni.princeton.edu}

\title{Three Simple Stokeslet Trajectories\footnote{Dedicated to William B. Russel, and especially his early work on suspension hydrodynamics}}
\keywords{suspension hydrodynamics}

\begin{document}

\begin{abstract}
  Exact results for three trajectories of small numbers of particles interacting hydrodynamically through Stokeslets are presented.
  First, the middle particle in a vertical trio of particles sediments at a constant velocity for all time.
  Second, a horizontal pair of particles sedimenting toward a rigid wall undergo a finite and surprisingly constant horizontal displacement in the limit of large initial separations.
  Third, a pair of particles sedimenting in a quadratic flow, such as that of a fluid pumped in the direction of gravity, oscillates in a periodic orbit and admits a Hamiltonian formulation.
\end{abstract}

\section{Introduction}

\begin{figure}[h]
  \centering
  \includegraphics[width=16cm]{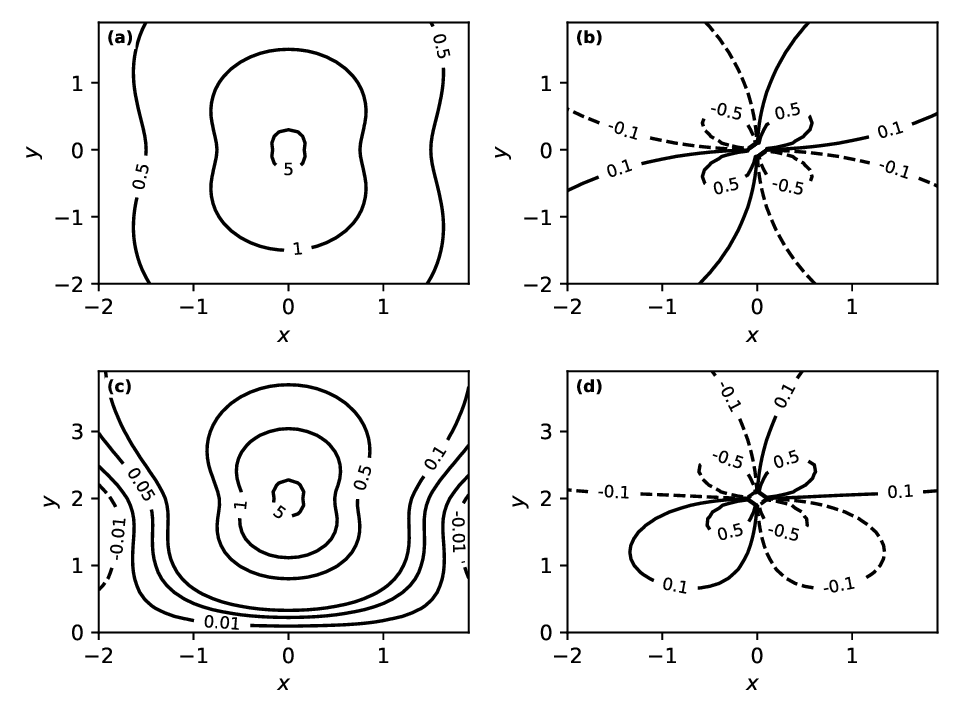}
  \caption{Particle-containing planes spanned by vectors parallel to ($y$ coordinates) and perpendicular to ($x$ coordinates) the applied force, which points in the positive $y$ direction here.
    (a) and (c) plot contours of constant $y$-component fluid velocity.
    (b) and (d) plot contours of constant $x$-component fluid velocity.
    (a) and (b) pertain to an infinitely-extended fluid.
    (c) and (d) pertain to a horizontal wall at $y=0$, where the no-slip condition applies.}\label{fig:velocity}
\end{figure}

A Stokeslet is simple, approximate model of hydrodynamic interactions between small particles driven by external forces at low Reynolds number and separated by far distances, i.e., much larger than a particle radius \cite{russel89}.
In an \textit{infinitely-extended} fluid, the disturbance velocity field at coordinate $\mathbf{x}$ induced by a forced particle at position $\mathbf{y}$ has a tensorial representation, which conveniently provides the disturbance field in response to a force $\mathbf{f}$ with arbitrary direction contracted with this tensor.
\begin{equation} \label{eq:stkfree}
  \mathbf{v}(\mathbf{x}, \mathbf{y}) = \frac{1}{8\pi\mu}\left[
    \frac{\mathbf{I}}{|\mathbf{x}-\mathbf{y}|}
    +
    \frac{(\mathbf{x}-\mathbf{y})(\mathbf{x}-\mathbf{y})}{|\mathbf{x}-\mathbf{y}|^3}
    \right]\cdot\mathbf{f}
\end{equation}
Above, $\mathbf{I}$ is the identity tensor.
$\mathbf{v}$ is proportional to $\mathbf{f}$ and non-isotropic.
The velocity field is larger in magnitude in the direction parallel to the force: Particles push and pull more strongly in front and behind them.
It also decays slowly with the distance $\mathbf{v}\sim|\mathbf{x}-\mathbf{y}|^{-1}$.
Disturbance velocities of this sort have been derived for other boundary conditions, such as near a single solid wall\cite{blake71}, between two walls \cite{liron76}, and in periodically replicated systems \cite{hasimoto59}.
See Fig.~\ref{fig:velocity} for contours of disturbance velocities relevant to this work.
Rather than being a mere approximation of flow around a forced sphere, it is a Green's function, able to construct more complex flows by superposition.

Along these lines, the fluid velocity induced by a system of forced particles can be also approximated by a superposition of Stokeslets, the end result more accurate if particles are widely separated.
Approximating the motion of the particles as the superposed fluid velocity at the particle center added to the Stokes velocities of those particles, i.e., \textit{particle} velocities in response to external forces only, we obtain a \textit{mobility matrix} describing the relationship between particle force and the disturbance velocity relative to some background flow \cite{guazzelli11a}.
\begin{equation}
  \begin{bmatrix}
    \mathbf{v}_1 \\ \mathbf{v}_2 \\ \mathbf{v}_3
  \end{bmatrix}
  =
  \begin{bmatrix}
    \mathbf{M}_{11} & \mathbf{M}_{12} & \mathbf{M}_{13} \\
    \mathbf{M}_{21} & \mathbf{M}_{22} & \mathbf{M}_{23} \\
    \mathbf{M}_{31} & \mathbf{M}_{32} & \mathbf{M}_{33}
  \end{bmatrix}
  \cdot
  \begin{bmatrix}
    \mathbf{f}_1 \\ \mathbf{f}_2 \\ \mathbf{f}_3
  \end{bmatrix}  
\end{equation}
In short, the velocity of a particle is proportional to the force on that particle and on other particles.
This is shown for three particles, which is the most particles considered in this paper, but it is generalizable to an arbitrary number of them.
Subscripts here identify particles.
The quantities $\mathbf{M}_{i\ne j}$, called \textit{pair mobilities}, or in the case where $i=j$ called \textit{self mobilities}, in general depend on coordinates and particle radii.
This simple formulation of particle hydrodynamics has already led to insight about sedimentation phenomena such as backflow \cite{saffman73}, suspension clouds \cite{ekiel-jezewska06}, and velocity fluctuations \cite{caflisch85}, which assumed simple distributions of Stokeslets rather than considering individual particle trajectories.

For small particle systems, with equal radii $a$ and identical forces $\mathbf{f}$, it is possible to derive exact results for the trajectories of the particles, which may be applicable to situations where particles are both widely separated, justifying the Stokeslet approximation, and further away from other particles in the same fluid.\footnote{Outside of special cases, systems with three or more Stokeslets are chaotic, presenting difficulties for this mathematical analysis \cite{janosi97}.}
The first such result, which has been known for a long time, is that a pair of identical particles in an infinitely-extended fluid sediments in a constant direction which depends on the particle separation, such that the particles neither come together nor drift apart.
Such results often have rheological interest.
For instance, although working with hydrodynamics more accurate than Stokeslets, \citeauthor{batchelor72a} derived the first correction with respect to particle concentration of the shear viscosity through the analysis of trajectories of two particles in a straining flow \cite{batchelor72a}.

\begin{figure}[h]
  \centering
  \includegraphics[width=16cm]{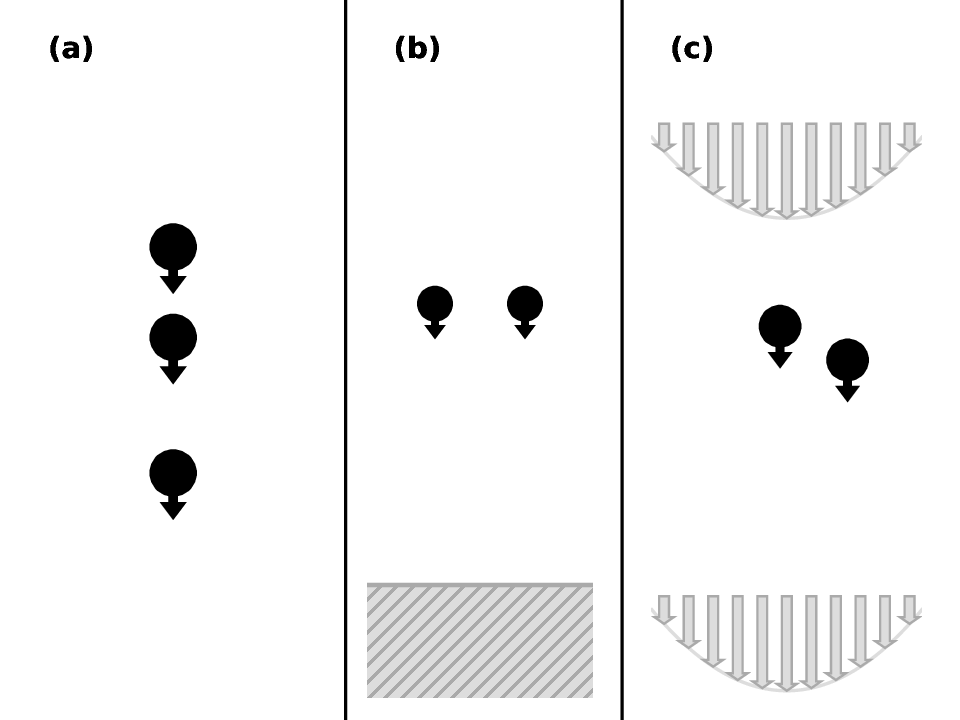}
  \caption{The three sedimentation scenarios examined in this work: (a) a vertical trio, (b) a horizontal pair above a wall, and (c) a particle pair in a quadratic flow.
    Black arrows extend from the identical spherical particles in the direction of the buoyant force.
    In (b), the rigid horizontal wall below the particles enforces a zero-velocity condition.
    In (c), velocity vectors of an example ambient quadratic flow are shown.
    (A quadratic flow oriented relative to the buoyant force as in this example induces oscillatory motion.)
    In all cases, the motion of the particles is a consequence of external forces and the hydrodynamic interactions they induce.}\label{fig:cartoon}
\end{figure}

In this work appear three exact results on the trajectories of systems of Stokeslets.
First, the center particle in a vertical trio particles oriented in the direction of gravity travels at a constant velocity for all time, explaining observations of more accurate hydrodynamic models from decades ago.
Second, particles in a horizontal pair sedimenting towards a rigid wall undergo a horizontal displacement that is surprisingly \textit{independent of their initial horizontal separation}, given enough time to sediment.
Third, a pair of particles sedimenting in a quadratic flow similar to that in a pipe pumping fluid downwards undergoes period oscillations and enjoys a Hamiltonian representation.
These are shown in Fig.~\ref{fig:cartoon}, (a) through (c), respectively.
The first two results rely on the precise inverse-distance decay of the hydrodynamic interactions.
The third may have implications for constitutive modeling of heterogeneous suspensions.
The first and third results may serve as classroom exercises for suspension hydrodynamics and ordinary differential equations.
The second is likely poorly suited for this, due to the complexity of the reflected hydrodynamic interactions.

\section{Results and discussion}

\subsection{Central particle velocity in a vertical trio}

First we consider three identical particles in a vertical line, with center-to-center displacements parallel to gravity, which exerts identical forces to the particles.
Let the positions of the particles be $x_1$, $x_2$, and $x_3$, counting in the direction of decreasing gravitational potential (i.e., towards the floor).
Assume that the particles are in an infinitely-extended fluid.
For simplicity here and in what follows, we work in distance units of particle radii $a$ and time units of Stokes times $6\pi\mu a^2/|\mathbf{f}|$, where $\mathbf{f}=4\pi a^3\Delta\rho\,\mathbf{g}/3$ is the buoyant force.
The latter is the time it takes for a lone particle to translate a distance unit.
This system of units absorbs the force magnitude into the time scale and uses force vectors of unit magnitude in the equations.
Under these circumstances, the self-mobility of each particle $i$ is $\mathbf{M}_{ii}=\mathbf{I}$, and the pair mobility of particles $i$ and $j$ uses the form of the Stokeslet in Eq.~\ref{eq:stkfree}.
\begin{equation}
  \mathbf{M}_{i\ne j} = \frac{3}{4}\left[
    \frac{\mathbf{I}}{|\mathbf{x}_i-\mathbf{x}_j|}
    +
    \frac{(\mathbf{x}_i-\mathbf{x}_j)(\mathbf{x}_i-\mathbf{x}_j)}{|\mathbf{x}_i-\mathbf{x}_j|^3}
    \right]
\end{equation}
The equation of motion of this particle system in its single spatial dimension is the following.
\begin{equation} \label{eq:three-particle-system}
  \begin{aligned}
    \frac{\mathrm{d}x_1}{\mathrm{d}t} = 1 + \frac{3}{2(x_2-x_1)} + \frac{3}{2(x_3-x_1)} \\
    \frac{\mathrm{d}x_2}{\mathrm{d}t} = 1 + \frac{3}{2(x_2-x_1)} + \frac{3}{2(x_3-x_2)} \\
    \frac{\mathrm{d}x_3}{\mathrm{d}t} = 1 + \frac{3}{2(x_3-x_1)} + \frac{3}{2(x_3-x_2)}
  \end{aligned}
\end{equation}
Because the spatial coordinates of interest are in the same direction as the force, the unit force is omitted in the equation above.
Lone particles translate at a velocity of one, but their velocity increases above that due to hydrodynamic effects of the other two particles.
It is simpler to analyze the center-to-center displacements instead of the centers themselves, since the system is translationally invariant.
Let $x_{23} = x_3-x_2$ and $x_{12} = x_2 - x_1$.
\begin{equation}
  \begin{aligned}
    \frac{\mathrm{d}x_{12}}{\mathrm{d}t} &= \frac{3}{2x_{23}} - \frac{3}{2(x_{12}+x_{23})} \\
    \frac{\mathrm{d}x_{23}}{\mathrm{d}t} &= \frac{3}{2(x_{12}+x_{23})} - \frac{3}{2x_{12}}
  \end{aligned}
\end{equation}
This pair of ordinary differential equations (ODEs) can be rewritten as a single differential equation in $x_{12}$ and $x_{23}$, omitting $t$, by the standard trick of ``solving for $\mathrm{d}t$'' in each equation and equating the resultant expressions in each.
\begin{equation}
  \begin{aligned}
    \frac{2}{3}\frac{\mathrm{d}x_{12}}{\frac{1}{x_{23}}-\frac{1}{x_{12}+x_{23}}} = \mathrm{d}t = 
    \frac{2}{3}\frac{\mathrm{d}x_{23}}{\frac{1}{x_{12}+x_{23}}-\frac{1}{x_{12}}}
  \end{aligned}
\end{equation}
After some algebra, we obtain a separable first-order ODE, which we can solve, revealing a constant of integration, a conserved parameter in the dynamical system of three particles.
\begin{equation}
  \begin{aligned}
    \frac{1}{x_{12}} + \frac{1}{x_{23}} = c
  \end{aligned}
\end{equation}
It is proportional to the interparticle contribution to particle 2's velocity in Eq.~\ref{eq:three-particle-system}, rendering that particle's velocity constant for all time.

The constancy of the velocity of the central particle is a non-trivial result.
It is not merely the result of symmetry arguments, but it depends on the particular decay power of the long-range hydrodynamic interactions.
One may verify this by considering particle 2's velocity with an arbitrary decay power $\alpha$ for the interparticle mobility and checking if it is a constant of integration.
\begin{equation}
  \begin{aligned}
    c_\alpha =
    \frac{1}{x_{12}^\alpha} + \frac{1}{x_{23}^\alpha}
  \end{aligned}
\end{equation}
Taking its first temporal derivative, and combining terms, we arrive at the following.
\begin{equation}
  \begin{aligned}
    \frac{\mathrm{d}c_\alpha}{\mathrm{d}t}
    &=
    -\alpha\frac{1}{x_{12}^{\alpha+1}}\frac{\mathrm{d}x_{12}}{\mathrm{d}t}
    -\alpha\frac{1}{x_{23}^{\alpha+1}}\frac{\mathrm{d}x_{23}}{\mathrm{d}t} \\
    &=
    -f\alpha\frac{1}{x_{12}^{\alpha+1}}
    \left[
      \frac{1}{k_\alpha x_{23}^\alpha}-\frac{1}{k_\alpha (x_{12}+x_{23})^\alpha}
      \right]
    -f\alpha\frac{1}{x_{23}^{\alpha+1}}\left[
      \frac{1}{k_\alpha(x_{12}+x_{23})^\alpha} - \frac{1}{k_\alpha x_{12}^\alpha}
      \right]
  \end{aligned}
\end{equation}
Above, we introduced $k_\alpha$, a coefficient that weights the reciprocal powers in the interparticle mobility, which we may take to depend upon the power of the decay of the hydrodynamic interaction.
For $\alpha=1$ we have $k_1=3/2$.
Combining the two terms, we arrive at the following.
\begin{equation}
  \begin{aligned}
    \frac{\mathrm{d}c_\alpha}{\mathrm{d}t}
    &=
    -\frac{f\alpha}{k_\alpha}\frac
    {x_{23}(x_{12}+x_{23})^\alpha - x_{23}^{\alpha+1} + x_{12}^{\alpha+1} - x_{12}(x_{12}+x_{23})^\alpha}
    {x_{12}^{\alpha+1}x_{23}^{\alpha+1}(x_{12}+x_{23})^\alpha} \\
    &=
    -\frac{f\alpha}{k_\alpha}\frac
    {(x_{23}-x_{12})(x_{12}+x_{23})^\alpha + x_{12}^{\alpha+1} - x_{23}^{\alpha+1}}
    {x_{12}^{\alpha+1}x_{23}^{\alpha+1}(x_{12}+x_{23})^\alpha}
  \end{aligned}
\end{equation}
One may verify that the right-hand side vanishes for $\alpha=1$.
Instead, for $\alpha=2$, we have the following.
\begin{equation}
  \begin{aligned}
    \frac{\mathrm{d}c_\alpha}{\mathrm{d}t}
    &=
    -\frac{2f}{k_2}
    \frac{(x_{23}-x_{12})(x_{12}+x_{23})^2 + x_{12}^3 - x_{23}^3}{x_{12}^3x_{23}^3(x_{12}+x_{23})^2}
  \end{aligned}
\end{equation}
The numerator is nonzero, so the quantity proportional to particle 2's velocity, and therefore the velocity itself, is not conserved for other power-law decays in pair interaction.

\citeauthor{leichtberg76} studied this three-particle problem with more accurate hydrodynamics through the discrete point boundary method \cite{leichtberg76}.
They noted the following about the near constancy of the velocity of particle 2 (emphasis mine).
\begin{displayquote}
  After the initial unsteady period (which is shrunk to zero at $\mathrm{Re}_\infty=0$, […]), spheres 1 and 2 possess essentially the same velocities, which are 30–40\% greater than the velocity of sphere 3.
  \textit{The velocity of sphere 2 does not vary greatly from this point to the end of the experiment.}
\end{displayquote}
With the more realistic hydrodynamics used by these authors, the velocity of particle 2 is not conserved, but the results in the present work finally give rationale for the near constancy they noticed.
It is due to the leading-order hydrodynamics and specifically the inverse-distance decay.

\subsection{Displacement of a horizontal pair above a wall}

Second we consider the sedimentation of a pair of identical particles arranged horizontally above a flat, rigid wall.
The Stokeslet pair mobility in this geometry can be constructed by a system of singularities at mirror-image points within the wall \cite{blake71,swan07}.
It consists of a inifinitely-extended-fluid Stokeslet of Eq.~\ref{eq:stkfree} and several ``reflections'' across the wall's surface: another Stokeslet, its gradient, and its Laplacian, with the latter two weighted by the distance from the forced particle to the wall.
The velocity fields used to construct it enforce the no-slip condition at the surface: zero velocity for an assumed stationary wall.
See Fig.~\ref{fig:velocity} (c) and (d) for level sets of this sum.
\begin{equation}
  \begin{aligned}
    \mathbf{M}_{\text{wall},i\ne j}(\mathbf{x}_i, \mathbf{x}_j)
    &= \mathbf{M}(\mathbf{x}_i, \mathbf{x}_j)
    -
    \mathbf{M}(\mathbf{x}_i, \mathbf{x}_j') \\
    &+ 2(\mathbf{x}_j\cdot\mathbf{e}_{\text{w}})
    (\mathbf{P}\cdot\boldsymbol{\nabla}_{\mathbf{x}_j'}\mathbf{M}(\mathbf{x}_i, \mathbf{x}_j')\cdot\mathbf{e}_{\text{w}})^\intercal
    +
    (\mathbf{x}_j\cdot\mathbf{e}_{\text{w}})^2
    (\nabla_{\mathbf{x}_j'}^2\mathbf{M})(\mathbf{x}_i, \mathbf{x}_j')\cdot\mathbf{P}
  \end{aligned}
\end{equation}
We introduced a unit normal that points away from the wall and into the fluid $\mathbf{e}_{\text{w}}$, a tensor $\mathbf{P}\equiv\mathbf{I}-2\mathbf{e}_{\text{w}}\mathbf{e}_{\text{w}}$ that reflects across the wall, and a primed symbol $\mathbf{x}_j'\equiv\mathbf{P}\cdot\mathbf{x}_j$ which is shorthand for the coordinate reflected from $\mathbf{x}_j$.
As written, ``target'' particle index $i$ is the particle receiving the velocity disturbance produced by force on ``source'' particle $j$.
Gradients are taken with respect to the reflected source particle's position $\mathbf{x}_j'$.
The corresponding self mobility replaces the first term with an identity tensor.
\begin{equation}
  \begin{aligned}
    \mathbf{M}_{\text{wall},ii}(\mathbf{x}_i)
    &= \mathbf{I}
    -
    \mathbf{M}(\mathbf{x}_i, \mathbf{x}_i') \\
    &+ 2(\mathbf{x}_i\cdot\mathbf{e}_{\text{w}})
    (\mathbf{P}\cdot\boldsymbol{\nabla}_{\mathbf{x}_i'}\mathbf{M}(\mathbf{x}_i, \mathbf{x}_i')\cdot\mathbf{e}_{\text{w}})^\intercal
    +
    (\mathbf{x}_i\cdot\mathbf{e}_{\text{w}})^2
    (\nabla_{\mathbf{x}_i'}^2\mathbf{M})(\mathbf{x}_i, \mathbf{x}_i')\cdot\mathbf{P}
  \end{aligned}
\end{equation}

When two identical particles with coordinates $\mathbf{x}_1(t)$ and $\mathbf{x}_2(t)$, with center-to-center separation vector $\mathbf{x}_2-\mathbf{x}_1$ perpendicular to $\mathbf{e}_{\text{w}}$, sediment towards a wall (i.e., gravity is parallel to $\mathbf{e}_w$), their distance from the wall $y(t)$ and their center-to-center separation distance $x(t)$ change simultaneously.
The ``un-reflected'' infinitely-extended-fluid part of the Stokeslet interaction drives the two particles to the wall, it reducing $y$, but it does not affect $x$.
The ``reflections'' embedded within the wall do, however, influence both $x$ and $y$, slowing the descent and pushing the particles apart.
Our objective is to consider the extent of the latter effect.
How far apart will particles spread apart from an initial center-to-center separation $L$?

For convenience, we consider the problem of sedimenting towards the wall in reverse, with particles falling from a wall instead, but the results apply when sedimenting towards the wall as long as certain quantities are negated.
This means that particles are driven together rather than pushed apart.
The linearity of the velocity with respect to the force means that our conclusions apply in the case of sedimentation towards the wall, as long as we suitably negate quantities.

After some algebra, the equations of motion for $x$ and $y$ are the following.
\begin{equation}
  \begin{aligned}
    \frac{\mathrm{d}x}{\mathrm{d}t} &= -18 \frac{x y^3}{(x^2 + 4y^2)^{5/2}} \\
    \frac{\mathrm{d}y}{\mathrm{d}t} &=
    1 - \frac{9}{8}\frac{1}{y} +
    \frac{3}{4}\left(
    \frac{1}{x}
    - \frac{1}{\sqrt{x^2 + 4y^2}}
    - 2\frac{y^2}{(x^2 + 4y^2)^{3/2}}
    - 24\frac{y^4}{(x^2 + 4y^2)^{5/2}}
    \right)
  \end{aligned}
\end{equation}
As-is, the set of equations is still difficult to analyze mathematically, and we seek to eliminate sub-dominant terms in the second equation to make things easier.
To make progress, we assume two conditions.
First, $x(t)$ does not change much from its large-enough initial value $L$.
Second, $1/y$ can be considered small enough compared to $1$, the leading term in the second equation.
We will justify our assumptions after our analysis by showing that our conclusions are consistent with them.
If $L$ is indeed large enough, the $1/x$ and $1/\sqrt{x^2+y^2}$ terms are clearly $\ll 1$ for all $y$.
So are the last two terms.
At intermediate and large $y$, relative to $L$, the terms are comparable to reciprocal powers of $L$, which are also small by assumption.
Thus, within our assumptions, at leading order we only need to retain the constant term in the second equation.
\begin{equation}
  \begin{aligned}
    \frac{\mathrm{d}x}{\mathrm{d}t} &= -18 \frac{x y^3}{(x^2 + 4y^2)^{5/2}} \\
    \frac{\mathrm{d}y}{\mathrm{d}t} &= 1
  \end{aligned}
\end{equation}
In other words, at leading order, the pair sediments towards the wall at the bare Stokes velocity, but hydrodynamic interaction from the wall still influences the center-to-center separation.
Again, we can solve for $\mathrm{d}t$ in both equations, leaving us with a single equation for the horizontal separation, a non-separable first-order ODE.
\begin{equation}
  \begin{aligned}
    \frac{\mathrm{d}x}{\mathrm{d}y}
    &=
    -18\frac{xy^3}{(x^2 + 4y^2)^{5/2}}
  \end{aligned}
\end{equation}
Now, we utilize our assumption that $|x(t)-L|\ll L$ and linearize $x$ about $x=L$, introducing a variable $\Delta\equiv x-L$ and retaining only leading terms.
This lets us integrate with respect to $y$ to obtain an estimate for $\Delta(y)$ in terms of $y$.
\begin{equation}
  \label{eq:delta}
  \begin{aligned}
    \frac{\mathrm{d}\Delta}{\mathrm{d}y}
    &=
    -18\frac{Ly^3}{(L^2 + 4y^2)^{5/2}} \\
    \Delta(y) &=
    \frac{3}{4}L\left[
      \frac{L^2 + 6y^2}{(L^2 + 4y^2)^{3/2}} - \frac{L^2 + 6y_0^2}{(L^2 + 4y_0^2)^{3/2}}
      \right] \\
    \Delta(u)
    &=
    \frac{3}{4}\left[
      \frac{1 + 6u^2}{(1 + 4u^2)^{3/2}} - \frac{1 + 6u_0^2}{(1 + 4u_0^2)^{3/2}}
      \right]
  \end{aligned}
\end{equation}
Above, we introduced the variables initial value $y_0$ to stand for the $y$ coordinate where $\Delta(y_0)=0$, $u\equiv y/L$ a scaled height, and $u_0\equiv y_0/L$ its initial value.
At $u\to\infty$, we have the ``final'' value of $\Delta$.
\begin{equation}
  \begin{aligned}
    \lim_{u\to\infty}\Delta(u)
    &=
    -\frac{3}{4}
    \frac{1 + 6u_0^2}{(1 + 4u_0^2)^{3/2}}
  \end{aligned}
\end{equation}
For convenience, we may select a $y_0$ large enough such that $1/y_0$ is smaller than some small error,\footnote{Choosing a large $y_0$ also avoids a known problem with the Stokeslet mobility near a wall, which causes particles very close to the wall to travel in the opposite direction of the force.} but then we may also choose an $L\gg y_0$ larger than this, such that the $u_0$ terms contribute nothing to the limit above, which holds in the limit of large $L$.
This justifies omission of the term proportional to $1/y$ in the dynamical equation for $y$.
By this procedure, we obtain our final result, which is clearly $\ll L$, consistent with our assumptions.
\begin{equation}
  \begin{aligned}
    \lim_{L\to\infty}
    \lim_{t\to\infty}\Delta(L, y_0, t)
    &=
    -\frac{3}{4}
  \end{aligned}
\end{equation}
This result, that the shrinking in horizontal separation of a widely-separated particle pair is both finite and largely independent of that separation, when given enough distance to travel away from the wall, is striking.
One might instead expect the change in separation $|\Delta|$ to diminish as the initial separation $L$ of the particles increases.
Despite this strangeness, the effect has a simple origin.
The integral performed in Eq.~\ref{eq:delta} is ``unitless.''
In other words, the integral is independent of the choice of units for length, but the left-hand side of the equation is a ratio of a center-to-center separation and the unit $a$.
Deeper than this, distances in the Green's function for Stokes flow near a solid boundary are coupled together in groups with units of inverse length, such that when they are integrated over distance become unitless.
Again, if the hydrodynamic interaction had a different spatial dependence, this result would not hold.

Now, we consider what happens when particles instead sediment towards rather than away from the wall.
Due to the linearity of Stokes flow, negating the forces is equivalent to rewinding the trajectory we just analyzed.
Therefore, the change in gap is just negated, and particles separate by three quarters of a radius.

It is worth considering how much headroom above a rigid wall a horizontal pair requires to realize most of this three quarters of a radius.
To achieve 99\% of this, assuming $u_0\ll 1$, we can solve $0.01 = (1+6u^2)/(1+4u^2)^{3/2}\approx 3/(4u)$ for $u$, and we find that $u\approx 75$, or a height of $75L$.\footnote{Lengths much larger than a particle radius must be compared against length scales where fluid inertia becomes as important as viscosity. We may estimate the latter by $a/Re$, where $Re$ is the Reynolds number \cite{guazzelli11a}.}

Finally, it must be mentioned that an infinite fluid above an infinitely-extending surface is an idealization.
It is also atypical that a pair of particles be perfectly identical, isolated from other particles, and arranged horizontally.
However, if the first two conditions can be guaranteed, deviation in horizontal orientation might not qualitatively change the result.
Generalizing to larger particle counts in a line or a plane is not straightforward, as particles in the center of the arrangement would sediment faster than particles on the perimeter, ruining the linear or planar arrangement.
If through an external potential the particles could be confined to a line or plane as they sediment, the configuration of the particles as they reach the wall might be interesting, especially if the line or plane of particles is initially disordered.
The balance between infinite range (independence of $L$) and bounded influence (finite $\Delta$) may produce interesting effects.
It should also be mentioned that in a real container that has walls on top and bottom, a pair of particles may contract together and then spread apart again as they fall from near the top wall to the bottom wall.

\subsection{Oscillation of a particle pair in a quadratic flow}

Third, we consider the sedimentation of another pair of identical particles, this time not restricted to a horizontal arrangement, through a specific ambient quadratic flow.
This is a quadratic flow where the flow's velocity is in the direction of gravity and its gradient is perpendicular to it.

Using the notation of \citeauthor{nadim91} \cite{nadim91}, we write the equation of motion as follows, but absorbing constant factors into the flow curvature $\mathbf{K}$ for simplicity.\footnote{The flow curvature $\mathbf{K}$ here, in units convenient to sedimentation, is equal to $6\pi a^3\mathbf{K}/|\mathbf{f}|$ when instead using the $\mathbf{K}$ of \citeauthor{nadim91}.}
\begin{equation}
  \begin{aligned}
    \frac{\mathrm{d}\mathbf{x}_1}{\mathrm{d}t}
    &=
    \mathbf{I}\cdot\mathbf{e}_{\text{f}}
    +
    \frac{3}{4}\left[\frac{\mathbf{I}}{|\mathbf{x}_{12}|} + \frac{\mathbf{x}_{12}\mathbf{x}_{12}}{|\mathbf{x}_{12}|^3}\right]\cdot\mathbf{e}_{\text{f}}
    +
    \mathbf{x}_1\mathbf{x}_1:\mathbf{K} \\
    \frac{\mathrm{d}\mathbf{x}_2}{\mathrm{d}t}
    &=
    \mathbf{I}\cdot\mathbf{e}_{\text{f}}
    +
    \frac{3}{4}\left[\frac{\mathbf{I}}{|\mathbf{x}_{12}|} + \frac{\mathbf{x}_{12}\mathbf{x}_{12}}{|\mathbf{x}_{12}|^3}\right]\cdot\mathbf{e}_{\text{f}}
    +
    \mathbf{x}_2\mathbf{x}_2:\mathbf{K}
  \end{aligned}
\end{equation}
The unit vector $\mathbf{e}_{\text{f}}$ points in the direction of the force, and the force magnitude is absorbed into the units.
We may introduce another unit vector $\mathbf{e}_{\text{h}}$ that points ``horizontally'' and perpendicular to the force.
For the problem at hand, the flow curvature $\mathbf{K}$ can be specified to $K\mathbf{e}_{\text{h}}\mathbf{e}_{\text{h}}\mathbf{e}_{\text{f}}$.
Neither the Stokesian drift of the particles nor the particular flow curvature we consider changes the center-to-center displacement $\mathbf{x}_{12}$ in the direction of $\mathbf{e}_{\text{h}}$.
Inserting the definition of the curvature and performing some algebra, we arrive at an equation of motion for the horizontal center $x_{\text{c}}\equiv(\mathbf{x}_1+\mathbf{x}_2)\cdot\mathbf{e}_{\text{h}}/2$ and the vertical displacement $\Delta y\equiv\mathbf{x}_{12}\cdot\mathbf{e}_{\text{f}}$, written in terms of the flow curvature scalar $K$ and the constant horizontal displacement $\Delta x$.

\begin{equation}
  \label{eq:smile}
  \begin{aligned}
    \frac{\mathrm{d}x_\mathrm{c}}{\mathrm{d}t}
    &=
    \frac{3}{4}\left[\frac{\Delta y\,\Delta x}{\left[(\Delta x)^2 + (\Delta y)^2\right]^{3/2}}\right] \\
    \frac{\mathrm{d}\,\Delta y}{\mathrm{d}t}
    &=
    K(x_2^2-x_1^2) \\
    &=
    2K\Delta x\,x_\mathrm{c}
  \end{aligned}
\end{equation}
We can also calculate the rate of change of the vertical center $y_{\text{c}}\equiv(\mathbf{x}_1+\mathbf{x}_2)\cdot\mathbf{e}_{\text{f}}/2$, but it is extraneous and can be expressed using the closed set of equations above.
\begin{equation}
  \begin{aligned}
    \frac{\mathrm{d}y_\mathrm{c}}{\mathrm{d}t}
    &=
    1
    +
    \frac{3}{4}\left[
      \frac{1}{\sqrt{(\Delta x)^2+(\Delta y)^2}} +
      \frac{(\Delta y)^2}{\left[(\Delta x)^2+(\Delta y)^2\right]^{3/2}}
      \right]
    +
    \frac{1}{2}K(x_1^2+x_2^2) \\
    &=
    1
    +
    \frac{3}{4}\left[
      \frac{1}{\sqrt{(\Delta x)^2+(\Delta y)^2}} +
      \frac{(\Delta y)^2}{\left[(\Delta x)^2+(\Delta y)^2\right]^{3/2}}
      \right]
    +
    K\left[x_\mathrm{c}^2 + \frac{1}{4}(\Delta x)^2\right]
  \end{aligned}
\end{equation}

As we did previously, we can solve for $\mathrm{d}t$ in both parts of Eq.~\ref{eq:smile} and equate the results.
The result is a constant of motion $H$ of the two-particle-in-quadratic-flow system.
\begin{equation}
  \label{eq:hamiltonian}
  H = K\Delta x\,x_\mathrm{c}^2 + \frac{3}{4}\frac{\Delta x}{\left[(\Delta x)^2 + (\Delta y)^2\right]^{1/2}}
\end{equation}
The symbol for this constant is chosen because the dynamical system has a Hamiltonian formulation
\begin{equation}
  \begin{aligned}
    \frac{\mathrm{d}q}{\mathrm{d}t} &= \frac{\partial H}{\partial p} \\
    \frac{\mathrm{d}p}{\mathrm{d}t} &= -\frac{\partial H}{\partial q}
  \end{aligned}
\end{equation}
where rates of change in ``position'' $q=\Delta y(t)$ and ``momentum'' $p=x_{\text{c}}(t)$ are written as suitable derivatives with respect to this Hamiltonian $H$.
\begin{equation}
  \begin{aligned}
    \frac{\mathrm{d}x_\mathrm{c}}{\mathrm{d}t}
    &=
    -\left[
      \frac{\partial H}{\partial\,\Delta y}
      \right]_{x_\mathrm{c}} \\
    &=
    \frac{3}{4}\frac{\Delta y\,\Delta x}{\left[(\Delta x)^2 + (\Delta y)^2\right]^{3/2}} \\
    \frac{\mathrm{d}\,\Delta y}{\mathrm{d}t}
    &=
    \left[\frac{\partial H}{\partial x_\mathrm{c}}\right]_{\Delta y} \\
    &= 2K\Delta x\,x_\mathrm{c}
  \end{aligned}
\end{equation}

The ``position,'' ``momentum,'' and ``energy'' in this formulation do not have units one would expect from mechanics.
In this formulation, the ``momentum'' has units of length, and the ``energy'' has units of squared length per unit time.
Hamiltonian formulations of suspension flows have appeared previously and have been used to explain the stability of horizontal polygonal arrangements of particles sedimenting through a quiescent fluid \cite{caflisch88}.
Such formulations are exceptional in that they may be used to prove boundedness of phase-space trajectories and establish periodic solutions, as in the case of a harmonic oscillator.
More on this later.

It is worth considering the linearized dynamical system, expanding in an assumed small parameter $\Delta y/\Delta x$, as this may reveal oscillation frequencies for certain values of the input parameters $x_{\text{c}}$, $\Delta x(0)$, $\Delta y(0)$, and $K$.
\begin{equation}
  \label{eq:expansion}
  \begin{aligned}
    H &= K\Delta x\,x_{\text{c}}^2 + \frac{3}{4}
    \left[
      1 - \frac{1}{2}\left(\frac{\Delta y}{\Delta x}\right)^2
      + \mathcal{O}\left(\left(\frac{\Delta y}{\Delta x}\right)^4\right)
      \right] \\
    \frac{\mathrm{d}x_{\text{c}}}{\mathrm{d}t} &= 
    \frac{3}{4\Delta x}
    \left[
      \left(\frac{\Delta y}{\Delta x}\right)
      + \mathcal{O}\left(\left(\frac{\Delta y}{\Delta x}\right)^3\right)
      \right]
  \end{aligned}
\end{equation}
The linearized system can be written in matrix form.
\begin{equation}
  \begin{aligned}
    \frac{\mathrm{d}}{\mathrm{d}t}
    \begin{bmatrix}
      x_{\text{c}}\\\Delta y
    \end{bmatrix}
    &=
    \begin{bmatrix}
      0 & \frac{3}{4\Delta x^2} \\
      2K\,\Delta x\ & 0
    \end{bmatrix}
    \begin{bmatrix}
      x_{\text{c}}\\\Delta y
    \end{bmatrix}
  \end{aligned}
\end{equation}
It has a constant solution $x_{\text{c}}=\Delta y=0$, which is also true for the full nonlinear system.
For $K=0$, the system has only a zero eigenvalue.
Otherwise, it has a pair of eigenvalues with opposite signs.
\begin{equation}
  \begin{aligned}
    \lambda &= \pm\sqrt{\frac{3K}{2\Delta x}}
  \end{aligned}
\end{equation}
If $K/\Delta x>0$, the eigenvalues are real, and the positive eigenvalue causes exponential departure from the neighborhood of the $x_{\text{c}}=\Delta y=0$ solution.
If $K/\Delta x<0$, the eigenvalues are imaginary, permitting small-amplitude oscillations of period $2\pi/|\lambda|$ about the $x_{\text{c}}=\Delta y=0$ solution.
The former, unstable case at is comparable to fluid in a vertical pipe being pumped upward while particles sediment downward.
The latter, stable case at has fluid pumped in the direction of gravity.

Without loss of generality, we assume that $\Delta x>0$ from this point onwards.
This makes $K>0$ map to unstable and $K<0$ to stable linearized scenarios.
It also sets the ranges of the Hamiltonian in the two cases.
For $K>0$, $H\in[0,+\infty)$, with the low end of the range pertaining to $x_{\text{c}}=0$ and $\Delta y=\pm\infty$.
For $K<0$, $H\in(-\infty,3/4]$, with the high end of the range pertaining to $x_{\text{c}}=0$ and $\Delta y=0$.

\begin{figure}[h]
  \centering
  \includegraphics[width=16cm]{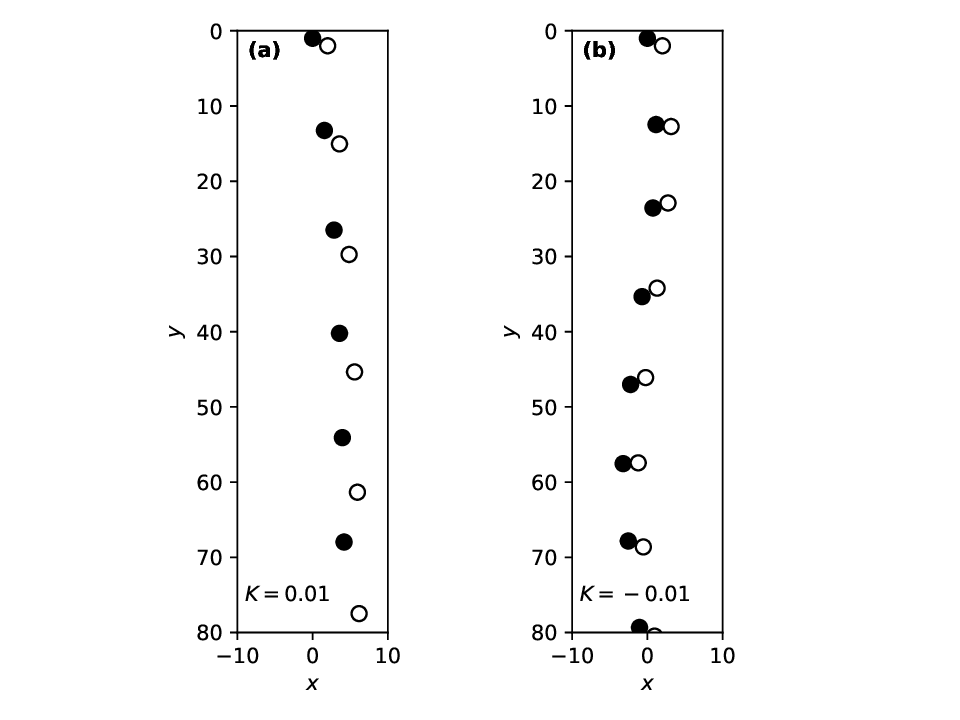}
  \caption{
    Pair trajectories for two different values of $K$.
    (a) shows non-periodic $K>0$ behavior.
    (b) shows periodic $K<0$ behavior.
    The force pushes particles in the positive $y$ direction.}\label{fig:trajectory}
\end{figure}

\begin{figure}[h]
  \centering
  \includegraphics[width=16cm]{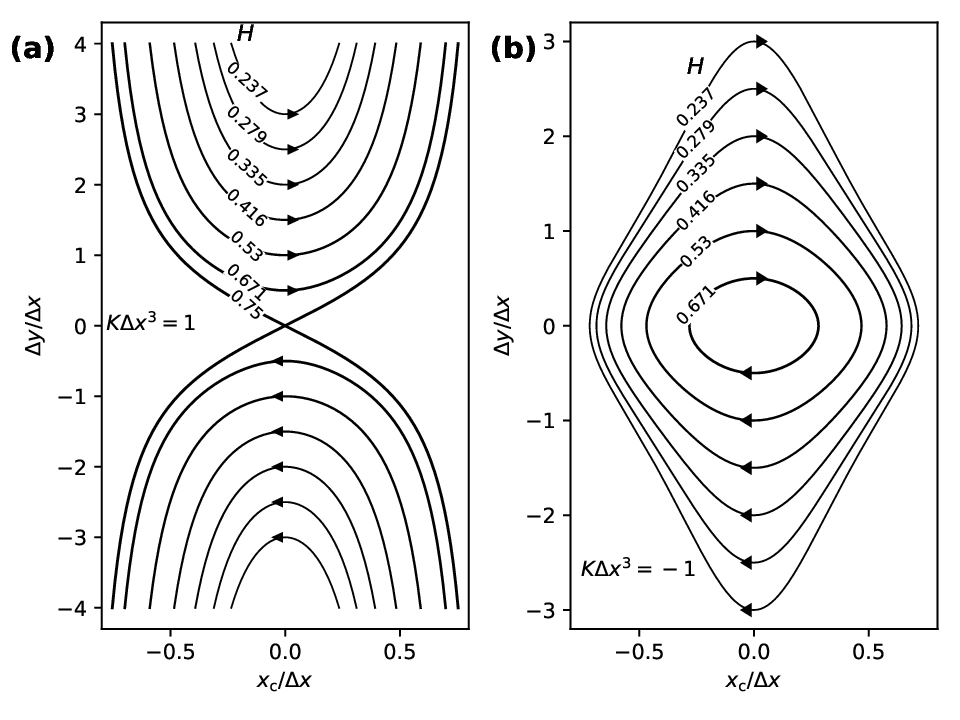}
  \caption{Phase portraits of the Stokeslet pair in a quadratic flow. (a) $K\Delta x^3=1$.
    (b) $K\Delta x^3=-1$. Values of $H$ decrease from the origin to the exterior.
    Arrows point in the direction of the system's temporal evolution.}\label{fig:phase}
\end{figure}

Now we consider full nonlinear solutions, i.e., avoiding the linearization of Eq.~\ref{eq:expansion}.
Example trajectories are sketched in Fig.~\ref{fig:trajectory}.
For $K<0$, like in the linearization, both $x_{\text{c}}$ and $\Delta y$ are bounded.
The bounds are provided by extremizing $x_{\text{c}}(\Delta y)$ or $\Delta y(x_\text{c})$ at constant $H$ (i.e., along a trajectory) in Eq.~\ref{eq:hamiltonian}.
$x_{\text{c}}^2$ is maximized when $\Delta y=0$, when it takes a value of $[(3/4)-H]/(-K\Delta x)$.
$\Delta y^2$ is maximized when $x_{\text{c}}=0$, where it takes a value of $\Delta x^2[9/(16H^2)-1]$.
Thus, the variables are always bounded along a trajectory.
The symmetry of the phase portrait on negating $x_{\text{c}}$ and $\Delta y$ establishes the closedness of the trajectories in phase space, so the solutions for $K<0$ are periodic.
Periodic $K<0$ and non-periodic $K>0$ trajectories in phase space are plotted in Fig.~\ref{fig:phase}.

The oscillations found in this problem bear resemblance to those found on sedimenting prolate ellipsoids in vertical pipes \cite{swaminathan06,huang14}, with obvious analogy between the Stokeslet pair and a prolate ellipsoid.
In those studies, the oscillations are created by mobility gradients and other hydrodynamic effects near the pipe walls.
It would be interesting to study the current oscillations in the presence of narrow tubes, and with more realistic hydrodynamics.
For the former, the image system introduced by \citeauthor{liron84} may be a good starting point \cite{liron84}.
For the latter, the approximation scheme used by \citeauthor{haber99}, which models the quadratic flow in the vicinity of a solid particle as a velocity gradient, may be helpful \cite{haber99}.

Such quadratic flows may be produced naturally, rather than being forced by flow through a pipe.
Suspension jets, or concentrated vertical assemblies of heavy particles, create an ensemble-averaged flow which is similar to the fluid-pumped-downwards stable case mentioned above \cite{pignatel09,crosby12}.
The suspension jets eventually break up due to a varicose instability into spherical suspension droplets, but it is worth considering the motion of pairs of particles inside those jets.
\citeauthor{crosby12} provides an axial velocity field $v_z(r)$ for a theoretical suspension jet in cylindrical coordinates, written in the frame where there is no net flux in the interior of the jet $r\le1$.
\begin{equation}
  v_z(r) = \begin{cases}
    \frac{1}{8} - \frac{r^2}{4} & \text{if }r\le1 \\
    -\frac{1}{2}\ln(r)-\frac{1}{8} & \text{if }r\ge1
  \end{cases}
\end{equation}
Velocity components in $\theta$ and $r$ directions are assumed to be zero.
The positive $z$ direction is the direction of sedimentation.
The second derivative of this velocity field in the radial direction is relevant for a particle pair with a component of center-to-center separation across the diameter of the jet, as the flow curvature it generates can cause the particle pair to oscillate back and forth rather than depart the jet.
\begin{equation}
  \frac{\partial^2 v_z(r)}{\partial r^2} = \begin{cases}
    -\frac{1}{2} & \text{if }r\le1 \\
    \frac{1}{2r^2} & \text{if }r\ge1
  \end{cases}
\end{equation}
This analysis assumes that the other particles in the jet act in a smeared-out fashion, rather than disturbing the particles in the pair, an assumption that should be interrogated.
The flow curvature then acts to localize the pair within the jet and eject the pair further out if they depart the jet, reflecting an affinity of particle pairs to the bulk of the suspension.

\begin{acknowledgement}
  The author thanks Roseanna N. Zia for introducing him to suspension hydrodynamics and for bringing related work to his attention.
  He also thanks Christopher W. Macosko for helpful suggestions on figures.
\end{acknowledgement}

\bibliography{bibliography}

\end{document}